# Beyond the Main Mode
The contribution of access and egress trips in door-to-door travel

**Nejc Geržinič**
Department of Transport and Planning, Delft University of Technology
n.gerzinic@tudelft.nl or nejcgerzinic@gmail.com

**Mark van Hagen**
NS Dutch Railways

**Hussein Al-Tamimi**
NS Dutch Railways

**Niels van Oort**
Department of Transport and Planning, Delft University of Technology

**Dorine Duives**
Department of Transport and Planning, Delft University of Technology

## Abstract

Access and egress trips constitute a substantial part of a train trip in minds of travellers, often being the deciding factor whether to travel by train at all. Despite a host of studies analysing individual legs within a multimodal trip chain, the full chain within a multimodal trip – including access, main and egress – has seen very limited attention. To understand the importance of all these choices, we use travel diaries from the Dutch Mobility Panel to estimate a nested logit discrete choice model. Our results suggest that as a main mode, train and bus/tram/metro (BTM) seem to be associated with an inherent disutility compared to walking, cycling or car. The in-vehicle time in train and BTM, however, seems to be perceived significantly less negatively (60% lower) than in private modes, making them comparatively more attractive for longer journeys. These results imply that, given the strong preference for walking for both access and egress, train stations should be sufficiently dense to allow most people to walk to a station. This, however, should not come at the expense of additional transfers, as they inflict substantial disutility. Operators need to find a balance between accessibility and directness. Given the strong dispreference of travelling by car to dense urban areas, these trips should be the primary target of policymakers and operators for attracting additional travellers to take the train. Future studies could further enhance our understanding of multimodal trips by including additional attributes in the data, account for respondent heterogeneity and study how individuals build their consideration set when making multimodal trips.

## 1 Introduction

Using the train is associated with additional trips, in order to get to and from the train station. These are referred to as access (before the train trip) and egress (after the train trip) trips or trip legs. According to research, these access and egress trips represent a substantial reason for the lower market share of train travel (Brons et al., 2009; Keijer & Rietveld, 2000). Krygsman et al. (2004) define the share of time spent on access and egress trips in relation to the total door-to-door travel time as the interconnectivity ratio. They show that for short trips (~30min), this ratio can be as high as 50%, and even for longer trips it is still 20-30%. Combining this with the fact that access and egress travel time tends to be valued more negatively (Wardman, 2004), the impact is even greater. Hoogendoorn-Lanser et al. (2006) suggest that despite making up less than 50% of the travel time, access and egress legs contribute around 2/3 of the total disutility to the train trip.

This suggests that a large part of the potential to shift travellers from car to train lies within the access and egress trips. Many researchers suggest that a better integration of cycling (and other micromobility) with the train could tap this potential (Martens, 2004, 2007; Rietveld, 2000). Jonkeren & Huang (2024) show that up to 5% of car trips and 10% of car distance travelled in the Netherlands could feasibly be replaced with a combination of the bicycle and PT, with train making up most of the PT part (>80%). And while these percentages may not seem substantial, they would constitute a doubling of the number of train trips.

Knowing that access/egress trips have a substantial impact on the choice to travel by train, while simultaneously also knowing there still is large, untapped potential in these multimodal trips, additional insights for the complete door-to-door trip may be necessary. Most studies thus far have focused primarily on the train (PT) trip and often only studying individual trip legs within the train trip (see Section 2). Our study adds to this body of knowledge by incorporating the various trip legs of a train trip and modelling the comprehensive door-to-door travel possibilities within a single model. The goal of this paper is to analyse the characteristics of (potential) train trips and include the information of all the individual trip legs in a single comprehensive model, giving us valuable information on not only the perception and weight of the individual legs, but also the relationships and trade-offs that travellers make between them when choosing how to travel.

The rest of the paper is structured as follows: a literature review of multimodal public transport trips and door-to-door travel preferences is presented in Section 2. The methodology adopted in this study is outlined in Section 3, with Section 4 providing a detailed description of the data processing required for the study. Section 5 then presents the main modelling outcomes, while Section 6 dives deeper into the results and analyses how individual attribute variation affects people's choices. Finally, the paper is summarised in Section 7.

## 2   State-of-the-art on multimodal trip chains

The importance of station accessibility has spurred a host of research on how access and egress travel are perceived. In this section, we provide an overview of studies that have analysed travel behaviour and traveller perception for different parts of the multimodal trip, including access/egress mode preference, travel time perception and stop/station choice. Some studies also analyse multiple of the listed aspects, i.e. a joint access-mode and station-choice study.

*Modal preference* - On the topic of modal preference, most studies agree that walking, cycling and bus/tram/metro (BTM) dominate train station access while walking and BTM are the modes of choice when egressing a train station (Bovy & Hoogendoorn-Lanser, 2005; Geržinič et al., 2025; Halldórsdóttir et al., 2017; Keijer & Rietveld, 2000; La Paix Puello & Geurs, 2014; Molin & Timmermans, 2010), with the specific context of the study influencing the exact order. Cycling tends to be a more popular means of access (and egress) in countries with more cycling culture, like the Netherlands (Bovy & Hoogendoorn-Lanser, 2005; Keijer & Rietveld, 2000; La Paix Puello & Geurs, 2014) and Denmark (Halldórsdóttir et al., 2017), whereas walking and even shared micromobility or ride-hailing are more common in places with a less developed cycling culture. (Azimi et al., 2021).

*Travel distance* - Comparing access trips for train versus BTM, Shelat et al. (2018) report that passengers are willing to travel further to access a train station than a BTM stop. Conversely however, the tipping point between choosing to walk or cycle is higher for BTM access, i.e. people would switch from walking to cycling at an access distance of 1.5km for a BTM stop but only 1km for a train station. The authors hypothesise that this could primarily be based on the bicycle parking facilities, which are much better at train stations, while being very minimal at BTM stops.

*In-vehicle time* - When considering the perception of in-vehicle time (IVT), the time spent accessing and egressing the stations tends to be perceived as more negative than time in the main mode. According to research, the scale of the main (train) IVT ranges from equal (Montes et al., 2023), to 50% lower (Geržinič et al., 2023), with most studies reporting a value of ~25%-30% below the access/egress IVT (Bovy & Hoogendoorn-Lanser, 2005; van Mil et al., 2021). Montes et al. (2023) report an equal IVT for all trip legs, but find vastly different perception of costs per leg, effectively resulting in a higher willingness-to-pay for travel time improvements on the access/egress legs. Geržinič et al. (2023) on the other hand report that the majority of travellers do not perceive train IVT differently from access IVT.

*Station choice* - Having carried out an extensive literature review on station choice, Young & Blainey (2018) find limited studies looking at mode-specific IVT on the access mode and outline this as a future research avenue. Bovy & Hoogendoorn-Lanser (2005) do report separate IVT perception, with private modes (walk, bicycle, car) being associated with more a negative IVT perception, whereas BTM and train IVT are almost indistinguishable. In contrast, La Paix Puello & Geurs (2014) find car IVT to be least negative, followed by BTM, cycling and walking as most negative. The same order, but then in terms of distance rather than time, is also concluded by Debrezion et al. (2009), although this is likely related to the average speeds of each mode, with the distance being more negative for slower-speed modes.

*Joint access-mode-and-station choice* - Access/egress mode choice is often taken together with the station choice. As summarised by Young & Blainey (2018), factors like train travel time, frequency, fare and number of transfers all influence station choice, which in turn influences access mode choice. Modelling this choice jointly shows that travellers are willing to make a longer access trip in exchange for improvements in the quality of the main leg. Travel time tends not to be a major factor, since train IVT is lower than access IVT, so any increase in access travel time would have to be compensated by substantially reduced train travel time. What most studies investigate is transfer avoidance; passengers may be willing to travel further, to a well-connected central station rather than their local station. Here too, the trade-offs differ between studies, with van Mil et al. (2021) and Yap et al. (2024) reporting a trade-off of only 5min of access time or 8min of train time to avoid a transfer. On the other hand, Geržinič et al. (2023) and De Bruyn et al. (2023) report a willingness to travel an additional 20-30min by train to avoid a transfer. Geržinič et al. (2023) also valued transfers against access time, showing that individuals are willing to travel 10-20min more, to a more distant station, if it means they have one transfer less on their train trip.

Although these studies investigate a wide array of attributes in multimodal train trips, most focus on a specific part of the trip, such as access/egress exclusively (Azimi et al., 2021; Halldórsdóttir et al., 2017; La Paix Puello & Geurs, 2014; Shelat et al., 2018; Ton et al., 2020), the joint access-mode-and-station choice (Bovy & Hoogendoorn-Lanser, 2005; Debrezion et al., 2009; Geržinič et al., 2023; van Mil et al., 2021), the access-main or main-egress mode choice (Molin & Timmermans, 2010; Montes et al., 2023) or purely on the train/PT trip (De Bruyn et al., 2023; Yap et al., 2024). To the best of our knowledge, no study has jointly considered the choices for access mode, boarding station, alighting station, egress mode and all of that in comparison with a private main mode, i.e. taking a car or bicycle instead of the multimodal train trip.

# 3 Methodology

To identify how much value travellers put on different aspects of a multimodal trip and how they stack up against a unimodal trip, we make use of a discrete choice modelling technique. Discrete choice models (DCMs) are a well-known and often applied method when quantifying the preferences and perceptions of people, with the goal of understanding what affects their choice behaviour. We start by defining the trips which we are interested in assessing in Section 3.1. Section 3.2 then elaborates on the data requirements for the DCM, followed by an explanation on how the choice models are specified in Section 3.3. A detailed description of the data used in the research is provided in Section 4.

## 3.1 Trips of interest

When analysing train trips and non-train trips which have mode-shift potential, a common approach is to analyse longer trips, typically above 5, 10 or even 15km (Jonkeren & Huang, 2024). While being a somewhat crude approach, studies do confirm that the train, especially in combination with micromobility (cycling), tends to only be a feasible alternative at longer distances, likely when the benefits of higher speeds offered by train travel can outweigh the substantial negatives experienced through access and egress legs (Hoogendoorn-Lanser et al., 2006). Jonkeren & Huang (2024) also confirm this to be a valid approach, with the average car trip being around 17km long, whereas trips showing mode-shift potential averaging in the range of 30-40km, with less than 1% of all bike-train trips being shorter than 8km.

That being said, train trips below these distances do occur and, even if not chosen, still provide valuable information on the decision making and trade-off behaviour of travellers. A different approach is to select trips which have a feasible train alternative, regardless of their length. After removing all trips below a distance of 10/15km, Jonkeren & Huang (2024) classified trips to be "potential train trips" if the door-to-door travel time was no longer than 1.5x that of the travel time by car (in some scenarios, ratios as low as 1 and 1.25 were also used). As their focus was bike-train trips, they also set a maximum access/egress distance for cycling.

We choose to apply a similar approach, while avoiding making assumptions as to how much people are willing to accept. Hence, we do not include an upper boundary on the travel time ratios for train trips, nor do we assume a maximum access/egress distance to/from a station. Instead, we apply a fairly simple algorithm: we check what is the closest train station to the origin and to the destination location. If this happens to be the same station, we conclude this not to be a feasible train trip. If the origin and destination are closer to different stations, we assume it to be a feasible train trip.

## 3.2 Choice set development

DCMs require choice observations, and the two typical data sources are either stated preference (SP) or revealed preference (RP) (Hensher et al., 2008; Train, 2009). SP data is obtained through surveys and is often used when analysing hypothetical future scenarios or options that either do not yet exist or are present on a very small scale. It also gives the researcher full control over the experiment. RP on the other hand relies on observations of behaviour, making the data more realistic, but requiring the researcher to make assumptions on the choice set of the decision-maker. Post-hoc assembly of choice sets is therefore a key part of RP data.

In this study, we use RP data to analyse the preferences of travellers for the entire door-to-door trip. In multimodal trips travellers have a series of choices to make, namely access mode, boarding station, alighting station and egress mode (see Figure 1). Constructing alternatives that contain all these choices in an SP survey is very difficult survey, as it would likely overwhelm the respondents with the number of questions in the survey. As outlined in Section 2, previous studies usually analysed one or two parts of the multimodal trip chain, in large part due to the complexity of the SP survey. Through the RP approach, all these choices are included in the data.

To construct the choice sets however, we need to make assumptions on the alternatives that may have been available and that the travellers had considered. In our study, we consider an alternative to be a complete door-to-door trip; this means that if only one element along the trip differs (different access mode, different home-end station used etc.) it constitutes a different alternative. Many different approaches exist on how to build a choice set, with an overview provided by Arriagada et al. (2025). They show that using past choices outperforms heuristic methods. However, due to the nature of our dataset, applying any form of past choice analysis to reconstruct the choice set is infeasible. This is because respondents only report three travel days in their diary, limiting the information we have at our disposal on their travel. Given this, we opt for the heuristic labelling approach where we select a handful of best objective alternatives, i.e. with the shortest travel time, the closest station etc. For train stations, we use the distance-based method (Young & Blainey, 2018), taking the nearest n stations to the origin and destination into the choice set. To balance choice set size and still accounting for realistic options among stations, we choose n=3. From our data, we see that 94% of all train trips used one of the three closest origin stations and 96% one of three closest destination stations. Increasing n to 4 or 5 would allow us to capture an additional percentage or two of train trips in our data, while drastically increasing the choice set size. An alternative would have been the catchment area method (all stations within a given radius), although this method is discouraged by Young & Blainey (2018). We choose not to use this approach as it can result in substantial differences in choice set sizes, further complicating what is already a complex model structure. It could result in some trips (in more rural areas) not having a single train station within the catchment area while other trips (in larger cities) having over 10 available stations. We continue to build up the choice set by including the best option (the ideal as provided by the routing API) of each access and egress mode for each of the selected stations, guaranteeing that each access mode is represented. This means we circumvent the issue of underrepresenting walking and cycling as access modes (Hoogendoorn-Lanser & Van Nes, 2004). We specify walking, cycling, bus/tram/metro (BTM) and car as access options, with the first three also included as egress options. Although car can also be an egress mode (taxi, car sharing, car passenger), our data included too few records of this to allow us to model it as a separate mode. For non-train trips, we also apply the labelling approach, finding the best option for each mode (walk, bicycle, car, BTM). We include a single unimodal alternative per mode as the dataset does not contain information on the route taken by the respondents. We are thus unable to account for route choice of the unimodal alternatives. Given all the possible combinations of access modes (4), boarding stations (3), alighting stations (3) and egress modes (3), along with the non-train alternatives (4), we end up with a choice set size of 112. The full overview of choices can be found in Figure 1.

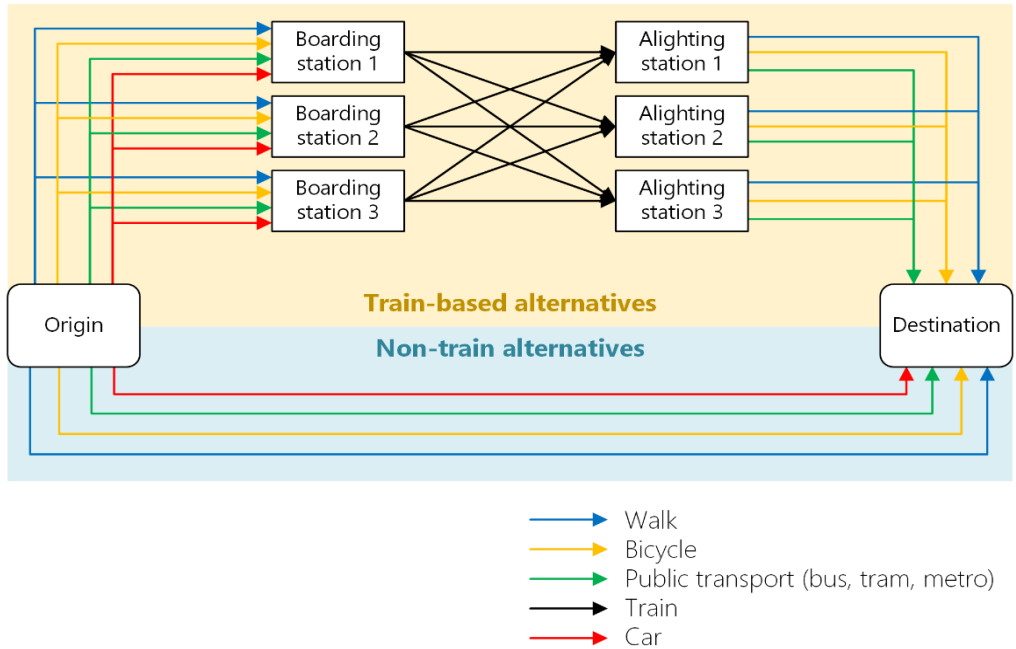

*Figure 1. Overview of all the different options that travellers have when making unimodal and multimodal train trips*

## 3.3 Modelling methodology

Choice data is then analysed by means of a DCM, where we assume that travellers make choices based on the random utility maximisation (RUM) decision rule (McFadden, 1974). Although other decision rules exist and there is no conclusive evidence that RUM-based models are superior (Chorus et al., 2014; Van Cranenburgh & Chorus, 2018), they are most often applied. Moreover, testing for different decision rules is beyond the scope of this study.

We estimate a series of multinomial logit (MNL) models. Although not optimal, they provide us with a good starting point in the model estimation process. Given that many choices in our data are correlated, i.e. choosing a bicycle+train combination likely shares characteristics with the walk+train combination, we also need to use a model that can account for such underlying structures. We choose to apply the nested logit (NL) model. While a mixed logit model can also account for nesting structures, and can also capture panel effects and respondent heterogeneity, the downside is that it is an open form model, requiring simulation and random draws, meaning that there is not a unique solution to the estimation. Given the scale of the modelling task, estimating an open form model which requires simulations was excluded due to its substantial computational cost, which exceeded the practical time and resource constraints of the present study.

NL on the other hand is closed form, making computationally much simpler (Hensher & Rose, 2007).

To apply the NL model, a nesting structure to model needs to be defined. The simplest option is to create a single nest with all alternatives with the train in that nest. This structure focuses on the distinction based on the main mode. Alternatively, nests can be made for each of the four choice components of the train trip, i.e. access mode, boarding station, alighting station or egress mode. Thirdly, based on the nesting structure proposed by van der Tuin et al. (2023), we also specify nests based on two of the four options; one nest is "walking to station A", then "cycling to station A" etc. This also results in four possible nesting structures, namely:

1. Access mode – boarding station (similarities based on the access leg)
2. Access mode – egress mode (similarities based on modal preferences)

3. Boarding station – alighting station (similarities based on train preferences)
4. Alighting station – egress mode (similarities based on the egress leg).

Other combinations are possible, for example "access mode – alighting station", but the connections between them are not logical.

## 4 Dataset description

The dataset used in this study features travel diaries from the Dutch Mobility Panel (MPN) (Hoogendoorn-Lanser et al., 2015). Panellists are asked to fill in a 3-day travel diary on a yearly basis, along with a series of socio-demographic questions to complement the travel information. We use data from the 2022 wave.

This following section provides additional information on how this data was filtered (Section 4.1), how the choice sets were constructed (Section 4.2) and finishes with descriptive statistics on the panellists' travel behaviour (Section 4.3).

### 4.1 Data processing

The first step of data preparation includes removing all erroneous and irrelevant data, as well as formatting the data for the following steps. The full dataset contains 38,808 records, which are filtered as follows:

1. **Erroneous data**: missing or incorrectly specified data (postcode, departure and arrival times, train station…) is removed.
2. **Tours**: trips where the origin and destination are the same, with no intermediate stop.
3. **Trips not originating at home**: trips which do not start at home have strong modal dependencies from the previous trip, i.e. if a person leaves home by car, they will likely undertake the following trips by car as well and will bring it back home. We do not make a distinction based on trip-chain complexity (how many trips are in a home-to-home tour), as Schneider et al. (2021) show there is no relation between trip-chain complexity and mode choice.
4. **Trips without a feasible train alternative**: as outlined in Section 3.1, we do not apply a minimal cut-off distance, nor do we assume a maximal ratio of travel time or access/egress distance (Jonkeren & Huang, 2024). Instead, we check for the nearest train station to the origin and destination. If station closest to both the origin and destination is the same, the trip is removed. If they are close to different stations, the trip is kept.

This filtering approach reduces the dataset to 5,276 feasible trips. Lastly, the trips data is unified and simplified:

1. **Aggregate mode coding**: the MPN records 29 possible modes. We merge this it into five mode groups: (1) walking, (2) cycling (all different types of bicycles, including folding, racing, electric…), (3) bus, tram or metro (BTM), (4) train and (5) car (different sizes, propulsion types and including both as a driver or passenger). A detailed overview of mode grouping and their frequency is provided in Table 4 in Appendix A.
2. **Remove walking trip legs during transfers**: due to the inconsistency of recording walking during transfers, we decide to remove all walking trip legs that occur as part of a transfer. Although walking during transfers is common and transfers can vary substantially with respect to walking distance and thus can have substantial impact on the choice to travel using PT, we remove walking legs between other modes, as respondents did not consistently report this. We do keep walking if recorded as the first access leg, the last egress leg or the main and only mode used for the trip.
3. **Merge successive trip legs with the same mode**: due to inconsistency of how transfers between the same mode are reported, we merge two successive trips with the same mode into one leg. Some respondents recorded the same mode twice in a row (i.e. bus and bus) to indicate having made a transfer, whereas others did not. To standardise the data, we consider a new trip leg only if a different mode is used. Information on transfers is still included in the choice model based on data obtained during the choice set construction step (see Section 4.2).

## 4.2 Choice set construction

Next, we assemble the choice sets. As outlined in Section 3, RP data typically does not contain choice set information. The researcher instead needs to make assumptions on what the consideration set was.

We use the labelling approach, which has the downside of being overcomplete with respect to the included alternatives, but guarantees that the choices are all represented (Arriagada et al., 2025). To obtain information on alternatives, we make use of the Google Directions API, Google Distance Matrix API and the Dutch Railways' own Reisinformatie (travel information) API.

As shown in Figure 1, we need information on both unimodal and multimodal travel alternatives. For unimodal trips for walking, cycling, and car, we make use of the Google Distance Matrix API, while for public transport, we use the Google Directions API. We use the origin and destination postcodes, which in the Netherlands represent a street. For **walking** and **cycling**, we extract the travel time and travel distance. For the **car**, we use the departure time and day of the week information from the dataset to obtain similar traffic conditions. We apply the departure time to be on the same weekday and time of day as recorded. To assess how much delay travellers may have experienced due to congestion, we also collect travel time data in free flow by using a departure time at 03.00. To capture other characteristics of the car, especially related to parking, we use a proxy for parking difficulty: urbanisation level. The assumption is that (in the Netherlands) more urbanised areas are more difficult to park in and cost more. For this, we use the data from the Central Bureau for Statistics (CBS) (CBS, 2024).

For **bus/tram/metro (BTM)**, we extract the in-vehicle time, walking time before and after the BTM trip (to the first stop and from the last stop) the number of transfers and modes used. We specify the preference for bus, tram and metro (not train), to specifically force the use of non-train alternatives. If no feasible alternative exists, the API will still suggest the train. In those instances, we consider the BTM alternative to be unavailable.

For the **train** trips, we (1) start by looking up the three closest stations to both the origin and destination (based on Euclidian distance). As outlined in Section 3, we limit ourselves to the nearest three to avoid exploding the size of the choice set. From the trips when a train was chosen, around 6% of trips used a boarding station and 4% an alighting station that was not among the three closest. In these instances, we code the used station as the $3^{rd}$, to retain this data. These trips always took place in larger urban areas, where passengers tend to have many easily accessible alternatives. Next, we (2) use the Reisinformatie API to extract the travel time, ticket price and the number of transfers for all nine travel combinations between the boarding and alighting stations. We use the API of the Dutch railways as the data is more up to date, includes price information, and the required transfer time is more realistic. In some cases, the same station may appear in both the access and egress set of stations. In that case, we consider that station pair to be unavailable.

For access and egress trips, we apply the Google Distance Matrix API and Google Directions API in the same way as for the unimodal trips for walking, bicycle, car and BTM, with a few minor differences:

1. For the car, we only consider the normal travel time at the stated departure time
2. For BTM, we do not force the use of bus, tram or metro, but let the API provide the most efficient solution. In cases where the train is offered, we consider that option irrelevant, as it indicates that the station is best accessible through other stations (usually one closer by), meaning we consider the BTM access/egress option to be unavailable.
3. For BTM, we only consider walking time at the start of the access trip or the end of the egress trip. Walking time during transfers is not considered, given the low reliability and variability of this data within the API.

Lastly, for price, only the Reisinformatie API contains price information, meaning we do not have direct data on the price of BTM or car, for either the access, main or egress legs. We thus impute these prices. For BTM, operators in the Netherlands use a based price + kilometre fare (GVB, 2025), which we combine with the trip distance obtained from the API. For the car, we use the average fuel price and average car efficiency, resulting in an approximate 0.1€/km fare, which we multiply with trip distance. Apart from a handful of tunnels, road pricing is not used in the Netherlands, so we do not need to consider the route. Lastly, bicycle parking facilities tend to be free, with a few that do charge only doing so after the first 24 hours (NS, 2025a); we thus do not consider any cost associated with parking one's own bicycle. For the bicycle on the egress leg, we use the €4.65 price of the Dutch railways bike sharing, which is present at most railway stations (NS, 2025b).

## 4.3 Descriptive statistics

To get a better idea of the data, we present various descriptive statistics about the types of trips recorded in the dataset. Starting with socio-demographics (Table 1), we see that in most categories, our sample is well representative of the Dutch >18 population. We see a slight underrepresentation of individuals above the age of 80. With employment status, given the unavailability of all data, we cannot judge on the representativeness among students and retirees, however we do observe a slight underrepresentation of the working population. With education level, we see a slight underrepresentation of middle educated individuals (those with secondary and vocational education) and an overrepresentation of higher educated individuals (having a university degree). However, both with employment status and highest obtained education level, we cannot draw full conclusions on representativeness due to a sizeable share of respondents preferring not to respond to these questions. The data for the Dutch population are obtained from the 2022 data provided by the Dutch bureau for statistics (CBS) (CBS, 2022).

*Table 1. Socio-demographic characteristics of the sample*

| Socio-demographic | Category | Count | Share | Population(CBS, 2022) |
|---|---|---|---|---|
| Gender | Female | 1,331 | 52% | 50.3% |
| | Male | 1,251 | 48% | 49.7% |
| Age | 18-29 | 446 | 17% | 18.9% |
| | 30-39 | 415 | 16% | 16.2% |
| | 40-49 | 416 | 16% | 14.6% |
| | 50-59 | 455 | 18% | 16.3% |
| | 60-69 | 444 | 17% | 15.6% |
| | 70-79 | 324 | 13% | 12.1% |
| | >80 | 82 | 3% | 6.4% |
| Employment status | Working | 1,413 | 55% | 64.3% * |
| | Retired | 515 | 20% | 35.7% * |
|               |           Student | 239 |  9% |       |
|               |             Other | 288 | 11% |       |
|               | Prefer not to say | 127 |  5% |       |
| Education level ** |              Low | 661 | 26% | 26.3% |
|               |            Middle | 857 | 33% | 41.4% |
|               |              High | 936 | 36% | 32.3% |
|               | Prefer not to say | 128 |  5% |       |

* For employment status on the national level, only the working and non-working is displayed. Non-working thus includes all individuals who are retired, studying and others.
** Low: no education, elementary education or incomplete secondary education
   Middle: complete secondary education and vocational education
   High: bachelor's or master's degree from a research university or university of applied sciences

Next, we assess the exhibited travel behaviour. We start by analysing the modal split and distance classes, Figure 2 shows the modal split and total number of trips per trip distance, based on the main mode (Euclidian distance). The number of short trips is rather small, due to the filtering applied, as described in Section 4.1. The modal splits per distance class are within expectation for the Netherlands, with short trips being dominated by active modes, the car becoming dominant at trips over 3km and the train appearing as a major contender for trips over 10km.

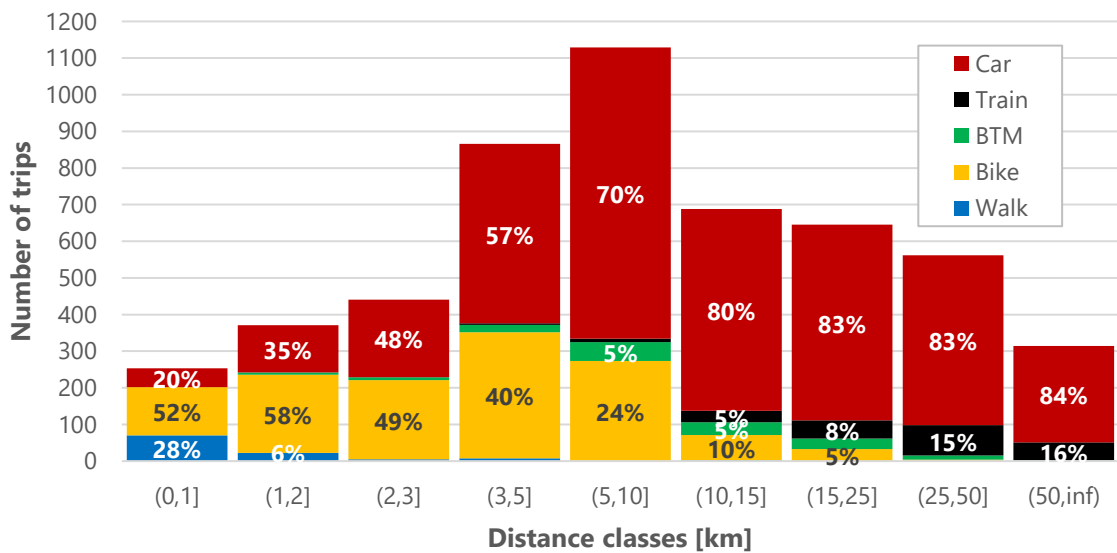

Figure 2. Distance distribution of recorded trips

Looking at trip lengths per mode in Figure 3, this trend reappears, with train trips being on average the longest (trips where train is the main mode) and active mode trips being on average the shortest. When the train trip is analysed per trip leg, walking and cycling have almost the same average distance on the access and egress side, indicating that both have an optimal/maximal length people are willing to travel. The averages of all access and egress trips reveal an important and known fact, that access trips tend to be longer. The difference for the car is due to different modes within the grouping "car", as shown in Table 4 in Appendix A; private car (as driver) is the dominant form on the access leg, whereas shared car passenger on the egress leg.

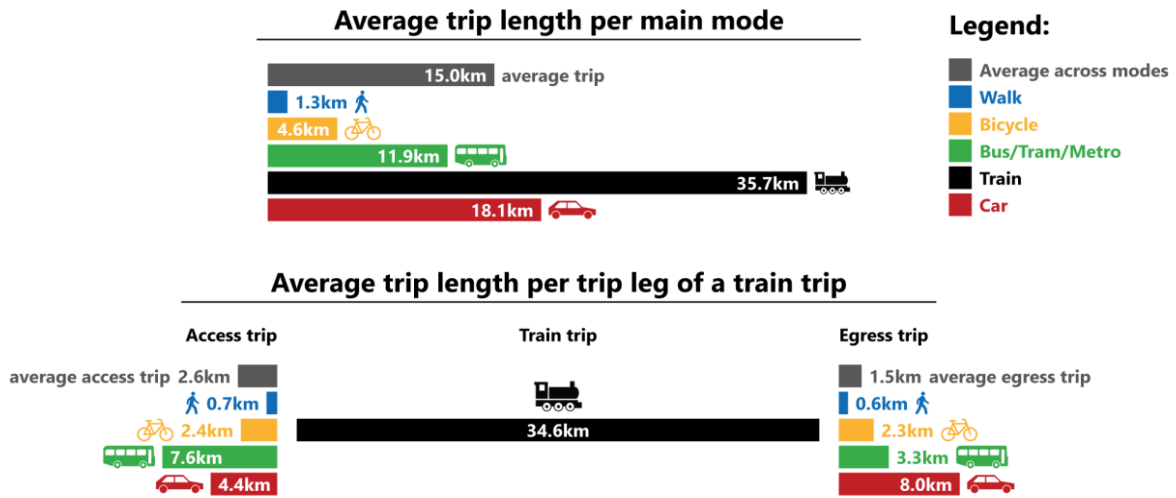

*Figure 3. Average trip lengths of different modes and trip legs[1]*

Looking at the combined choices of access mode, boarding station, alighting station and egress mode (Figure 4), we see that around half of all trips begin with a bicycle, with around a third being walking trips. On the egress leg however, walking dominates as an egress mode with two out of every three train trips ending with walking, followed by almost 20% of egress trips undertaken with BTM. The car comes in last both as an access or egress mode. The patterns of which modes are used to access which station are largely similar on both the access and egress side, with a large majority (>75%) of travellers use the closest stations for their train trips. Walking is almost exclusively used to access/egress the closest station, with cycling and car also primarily being used to travel to/from the closest station. BTM on the other hand tends to be used more often to access/egress more distant stations. This may imply that when travelling with BTM, the station is often not the closest, but the one best connected in the railway network, offering both better BTM connections and higher quality train services.

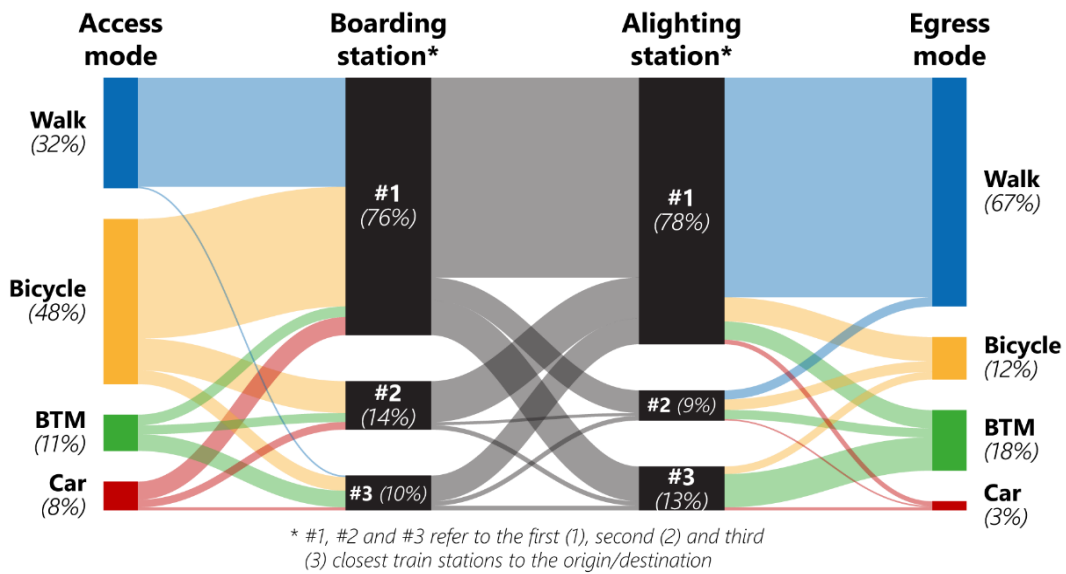

*Figure 4. Access and egress mode; and boarding and alighting station distribution in the data*

---

[1] The train trip distance between the two charts in Figure 3 differs. The first shows the average door-to-door trip distance where the train is the main mode. The second shows the average station-to-station distance of the train trip.

# 5  Modelling results

This section presents the final model specifications and parameter estimates, as shown in Table 2. The model was estimated using Pandas Biogeme (Bierlaire, 2023). For each group of parameters, the various tested specifications are discussed. The final model achieved a model fit of 0.8459 adjusted rho-square and a BIC value of 7,095.

Testing a wide array of nesting structures, the simplest single-nest structure – all alternatives making use of the train – outperformed all others, both in terms of model fit and BIC value. The detailed modelling outcomes are presented in Table 5 in Appendix B. The nesting parameter was estimated to be 2.45, which translates into a correlation of 0.83 between the alternatives within the same nest. This signifies that modes within the nest primarily compete with each other and less with alternatives outside the nest. A new train-based alternative is therefore more likely to attract existing train travellers (but using different access/egress modes or stations) than non-train travellers, meaning the modal shift to the train is limited.

With respect to taste parameters, we note that no cost parameter is included in the final model. Despite including cost in a variety of approaches, including (1) as a separate parameter, (2) with value-of-time and (3) a time/cost ratio, all the resulting models were either not significantly better, the cost parameter was illogical (positive) or insignificant or all of the above. This issue arises because price was obtained through distance, meaning price and time are highly correlated, with a correlation coefficient of 0.97, meaning that price is likely already captured in the travel time parameter. We therefore decide to proceed with modelling without the cost parameter.

*Table 2. Nested model parameter estimates, and model fit*

| | | | | |
|---|---|---|---|---|
| Parameters | | 24 | | |
| Null Loglikelihood | | -22,514 | | |
| Final Loglikelihood | | -3,444 | | |
| Adjusted Rho-square | | 0.8459 | | |
| BIC | | 7,095 | | |
| | | | Estimate | t-val. |
| Nesting parameter | | μ | 2.450 | 9.03 ** |
| | | | | |
| Trip-related parameters | Mode | | | |
| Travel time | Walk, Bicycle, Car | | -0.082 | -13.50 ** |
| [min] | BTM, Train | | -0.034 | -7.91 ** |
| | Car traffic | | -0.015 | -0.94 |
| Transfer [constant] | BTM, Train | | -0.864 | -7.69 ** |
| Highly urbanised origin [constant] | Car | | -0.299 | -3.20 ** |
| Highly urbanised destination [constant] | Car | | -0.945 | -10.90 ** |
| | | | | |
| Mode-specific constants | Mode | | | |
| Main mode | Walk | | *0 (ref)* | |
| [constant] | Bicycle | | -0.140 | -0.69 |
| *Baseline trip purpose: Leisure* | Car | | -0.106 | -0.43 |
| *Baseline car availability: None* | BTM | | -1.850 | -7.29 ** |
| | Train | | -1.760 | -6.92 ** |
| Main mode | Walk | | *0 (ref)* | |
| [interaction] *Work trip* | | Bicycle | 2.130 | 4.09 ** |
| | | Car | 1.060 | 2.02 * |
| | | BTM | 2.150 | 3.92 ** |
| | | Train | 2.450 | 4.49 ** |
| Main mode [interaction] *Car available* | | Walk | 0 (ref) | |
| | | Bicycle | -0.154 | -0.70 |
| | | Car | 0.569 | 2.53 * |
| | | BTM | -0.961 | -3.51 ** |
| | | Train | -0.522 | -1.94 * |
| Access mode [constant] | | Walk | 0 (ref) | |
| | | Bicycle | -1.020 | -10.10 ** |
| | | Car | -2.050 | -10.40 ** |
| | | BTM | -1.810 | -9.69 ** |
| Egress mode [constant] | | Walk | 0 (ref) | |
| | | Bicycle | -1.860 | -10.70 ** |
| | | BTM | -1.750 | -11.10 ** |

*** p < 0.01, * p < 0.1*

## 5.1 Trip-related parameters

Starting with travel time, we test mode-specific and leg-specific travel time parameters to analyse if the perception differs per mode (walk, bicycle, car, BTM and train) and trip leg (access, main, egress). The results indicate that the perception of travel time does not vary between different trip legs, i.e. the same mode was always perceived the same no matter during which leg of the trip. Mode-specific perception shows that time spent walking, cycling or driving was not significantly different, whereas time spent in any PT mode (bus, tram, metro or train) was also not significantly different from each other, while being less negative than time walking, cycling and driving. Specifically, time in PT is only 40% as negative as time walking, cycling or driving. The most likely reason for this is that one has the freedom to do other things while in PT, making time less important. In other modes, one actively needs to drive/steer, pay attention to one's surroundings and cannot (should not) do anything else. Alternatively, this could also be related to the missing cost data due to issues with high collinearity.

For car trips, we also test if the additional travel time incurred due to traffic has a more negative perception and our results show it is not, with the parameter being insignificant. This suggests that while people do prefer to avoid traffic, purely for the sake of saving travel time and not necessarily that it is time that is perceived more negatively than regular driving time.

Lastly, we analyse the perception of transfers in PT legs. Here too, we test if transfers on a BTM trip or on a train trip differ, and the results suggest it does not. According to our results, a transfer is thus perceived equally in any PT mode. The penalty for a transfer is quite substantial, suggesting that travellers are willing to walk/cycle/drive to a station up to 10min farther, if it allows them to avoid one transfer. From the perspective of PT in-vehicle time, this means travellers are willing to ride a full 25min more if this enables them to avoid a transfer.

## 5.2 Alternative specific constants

Moving to mode-specific constants, we separate them into three separate groups for the different trip legs, namely access, main and egress, with walking used as the baseline in all three groups. Interaction effects only had a significant effect on mode preference in the main leg, with access- and egress-leg interactions producing limited and often insignificant impacts. The main leg preference is interacted with trip purpose (work or leisure) and if the respondent owned a car or not[2]. Figure 4 shows the scale of the modal preferences under different circumstances. We see that in general, travellers prefer private modes like the bicycle and car for work-related trips. As can also be expected, if the individual does not have a car, the preference for car is lower compared to other modes, although the difference is not substantial. It is also interesting to note that public transport modes are less preferred for leisure trips and more for work-related trips. This could be because the more regular travel patterns related to commuting, travel reimbursement for public transport and the fact that many offices are located in well accessible locations.

To mimic the difficulty of driving in more dense urban areas and finding parking there, we interact the five urbanisation levels with the car. Only the highest urbanisation level category achieved a significant interaction and as expected, it discourages driving in relation to other modes.

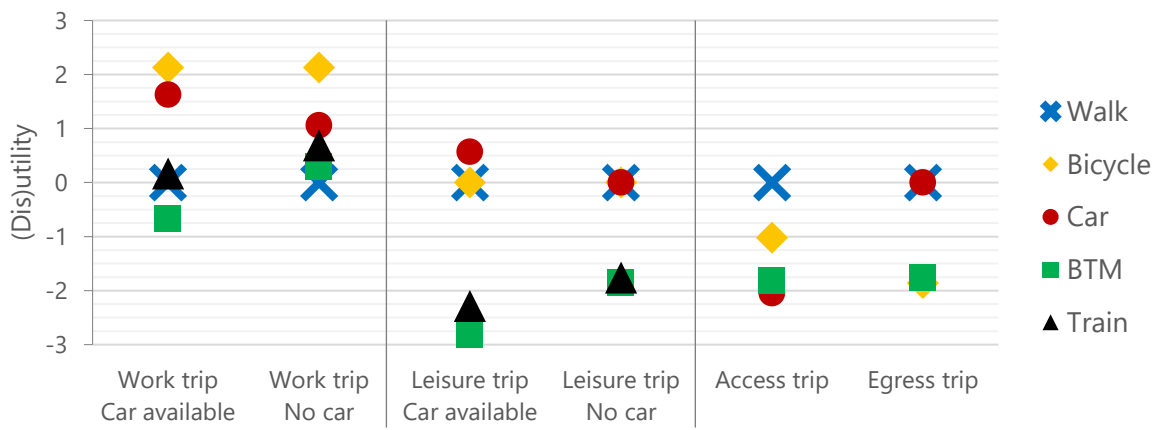

*Figure 5. Mode preferences under different context conditions and for different trips*

For both access and egress trips to/from the train station, walking is the most preferred mode (Figure 5). On the access side, it is followed by cycling. In the Netherlands, train stations have extensive parking facilities, with many towns having big, covered and supervising bicycle garages, making parking one's bicycle at the station very convenient. BTM and car come in third and fourth respectively. BTM connections to train stations tend to be good in the Netherlands, although the additional cost and coordinating train and BTM schedules likely makes it less appealing. On the other hand, the car offers more freedom, but parking at many train stations can be challenging. More rural stations often have park-and-ride facilities, but in towns this is typically not the case, making parking very difficult.

On the egress side, BTM and the bicycle are roughly equally (dis)liked, with a slight preference for the former. The comparatively lower preference for the bicycle on the egress trip compared to the access trip is likely since most travellers do not have a second bicycle available at the destination side, meaning they rely on shared bicycles if they wish to travel the last mile by bicycle.

# 6 Implications

As the impact and interactions of the taste parameters are not always immediately clear, this section provides additional insight into the trade-offs between parameters and what affects mode choice under different circumstances. The section starts by diving into the implications of the main-leg parameters, followed by access-leg parameters in Section 6.2 and egress-leg parameters in Section 6.3. Lastly, based on the first three sections, the policy recommendations are discussed in Section 6.4.

---

[2] Even if travellers do not own a car, we still include the car alternative, given that the "car" alternative includes also modes such as car passenger, shared car or taxi. The car can also be borrowed by someone in the household, as the car ownership data is recorded on an individual basis, not household basis.

## 6.1 Implications of the main leg

The main leg of the trip is characterised by mode-specific constants, two in-vehicle time (IVT) parameters, transfer and interaction effects. Starting with the comparison trade-offs between the IVT parameters and other constants, we show how much walk/bicycle/car (private mode) IVT and PT IVT travellers are willing to trade in Table 3.

We see that for example that if the train leg has a transfer, the travel time of private modes can be up to 10min longer to still be equally attractive. It also means that on the access or egress leg, people are willing to travel to a station up to 10min farther to avoid a transfer. For the car, highly urbanised regions are less attractive; specifically, a trip originating in a dense urban area equals a travel time penalty of ~3.6min, meaning for example that travellers accept a 3.6min longer cycling or walking time as opposed to less dense areas. For the destination, the penalty is higher at over 11min.

The same trade-offs can be seen for PT IVT, being are roughly 2.5x higher as PT IVT is seen as less negative, meaning that travellers are willing to accept more travel time in PT for making these same trade-offs. To avoid a transfer, people seem to be willing to travel some 25min longer. And for travelling to a dense urban area, the travel time can be 28min longer than a trip to a less dense area for it to remain equally appealing.

*Table 3. Trade-offs between different taste parameters and how they relate to perceived travel time*

|  | Walk, Bicycle, Car IVT [min] | PT IVT [min] |
|---|---|---|
| Walk, Bicycle, Car IVT [min] | - | 2.4 |
| PT IVT [min] | 0.4 | - |
| Transfer | 10.5 | 25.6 |
| Highly urbanised origin | 3.6 | 8.9 |
| Highly urbanised destination | 11.5 | 28.0 |

## 6.2 Implications of the access leg

As station accessibility is essential for its attractiveness, we start by analysing the impact of the availability of different modes to reach a train station. Faster modes allow for a larger catchment area, providing an acceptable level of accessibility for a larger number of people. Figure 6 (left) shows how the share of train trips would change based on how far the station is from someone's origin and which modes are available to them. The details of these calculations are explained in Appendix C. Figure 6 (right) then shows how much the share of train increases if the bicycle, BTM or car is added to an otherwise walk-only option. Both figures show that cycling seems to have by far the largest impact on improving station accessibility, having the potential to increase train market share by over 10 percentage points (p.p.) for certain access distances. BTM is primarily interesting for slightly longer access distances (>5km) compared to the bicycle, as it involves walking and waiting time, which can only be "worth it" beyond a certain travel time. Interestingly, car seems to impact the train market share the least, with limited benefits for all distances below 10km.

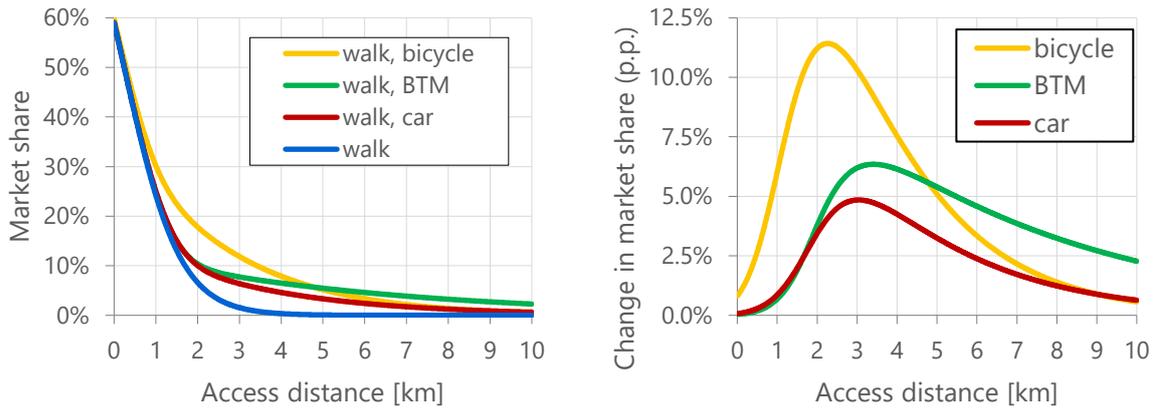

*Figure 6. Market share (left) and the change of market share (right) of train as the main mode, if bicycle, car or BTM are added as access modes, when walking is already possible. The change on the right is compared to the benchmark where walking is the only alternative.*

As BTM and car are likely to have some additional benefits for longer trips compared to cycling, we also analyse how their inclusion changes the attractiveness of a train station to which travellers can both walk and cycle. Figure 7 (left) shows that the increase in market share is not easily noticeable. Figure 7 (right) highlights that, as expected, having both BTM and car alongside walking and cycling provides the biggest increase in train market share. However, on their own, BTM has a much higher added value, increasing train market share by an additional 2 p.p., whereas the car only reaches a ~0.75 p.p. increase at most. BTM also increases the market share more at longer distances, which better complements the shorter distances reachable by walking and cycling.

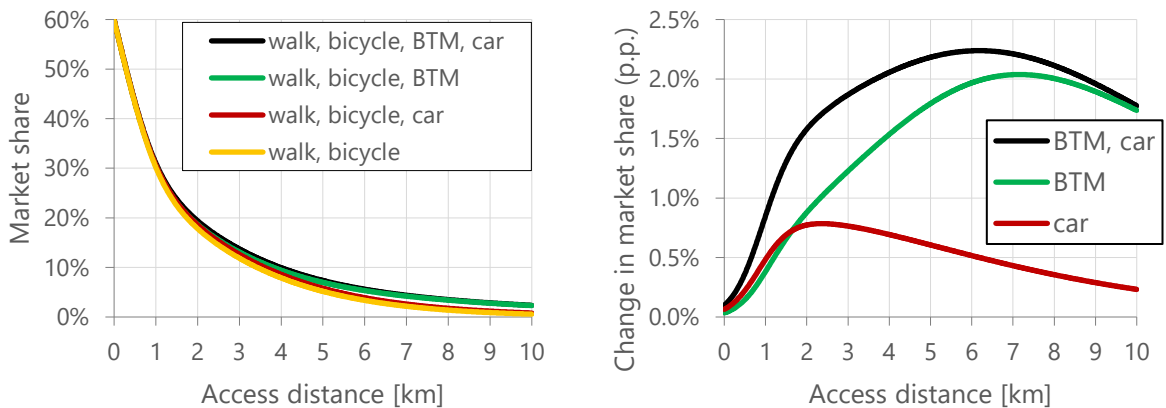

*Figure 7. Market share (left) and the change of market share (right) of train as the main mode, if car, BTM or both are added as access modes, when walking and cycling are already possible*

Lastly, we look at the modal split on the access leg on its own and focus on the impact of the quality of the BTM service, specifically walking time and transfers. The former can be seen as a proxy for the network density, as individuals need to walk further if the network is less dense. Below, we analyse the impact of a transfer, which has the same impact as an additional 10min walking time, which is also not unheard of. As shown in Figure 8, a transfer in the BTM leg can have a substantial impact on the modal split. In the case of a direct BTM trip (Figure 8 left), it becomes the dominant mode for all distances beyond 5km, taking over from the bicycle, while the car never surpasses a market share of 15%. If the BTM trip does require a transfer, the bicycle is the preferred mode of choice for access distances up to 8km, while the car's market share reaches almost 30% and is preferred to BTM up to 7km.

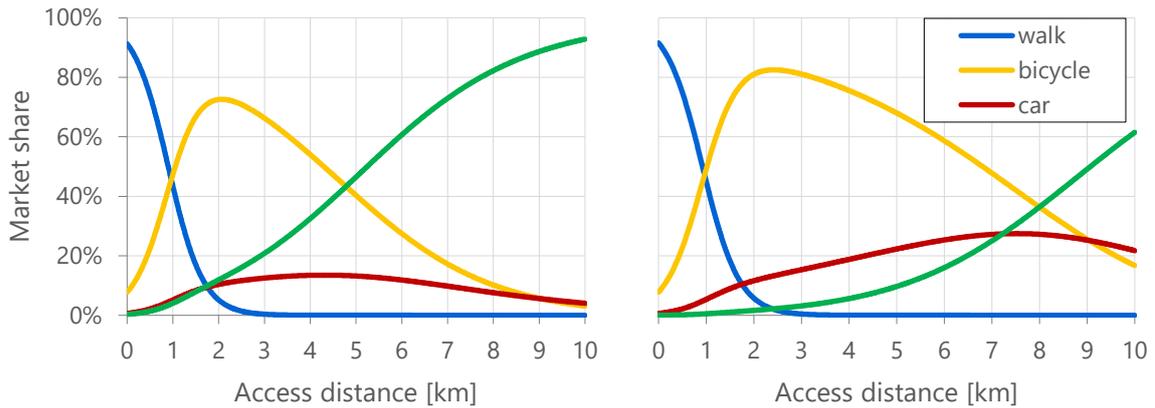

*Figure 8. Mode preferences on the access leg, if the BTM option has 0 transfers (left) or 1 transfer (right) for reaching the train station*

## 6.3 Implications of the egress leg

A similar pattern to what we see on the access leg can be observed on the egress leg. As seen in Figure 9, adding faster egress modes will often substantially increase the attractiveness of the train for distances of over 2km. Conversely to the access side however, BTM is the mode of choice on the egress leg. It's lower constant, combined with a lower IVT perception makes it much more attractive. On its own, it can attract double as many passengers as the bicycle (5 p.p. increase vs. 2.5 p.p. increase). The peak increase also occurs about 1km farther from the station. Including both modes has the biggest benefit, but interestingly, the addition of the (shared) bicycle, if (higher quality) BTM is already present, does not add that much added value for the train as the main mode.

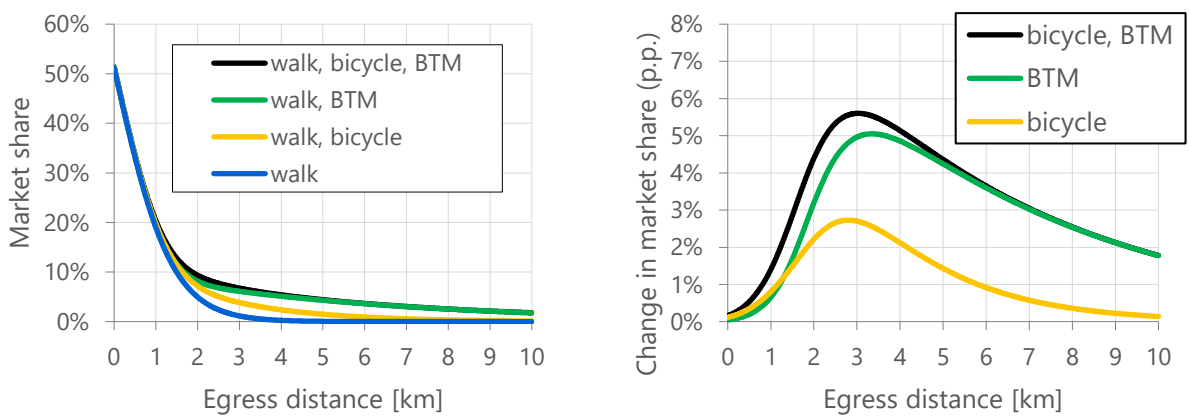

*Figure 9. Market share (left) and the change of market share (right) of train as the main mode, if bicycle, BTM or both are added as egress modes, when walking is already possible*

If travellers are required to transfer during their BTM egress trip however, the modal split changes drastically (Figure 10). With a direct BTM trip, it becomes the mode of choice already at distances beyond 1.7km, with the (shared) bicycle never becoming the most popular mode. With one transfer, the (shared) bicycle becomes substantially more appealing, surpassing walking at 1.8km of egress and being the preferred egress mode up ~4km, when the additional time and effort of cycling no longer outweighs the disutility of the transfer in the BTM trip.

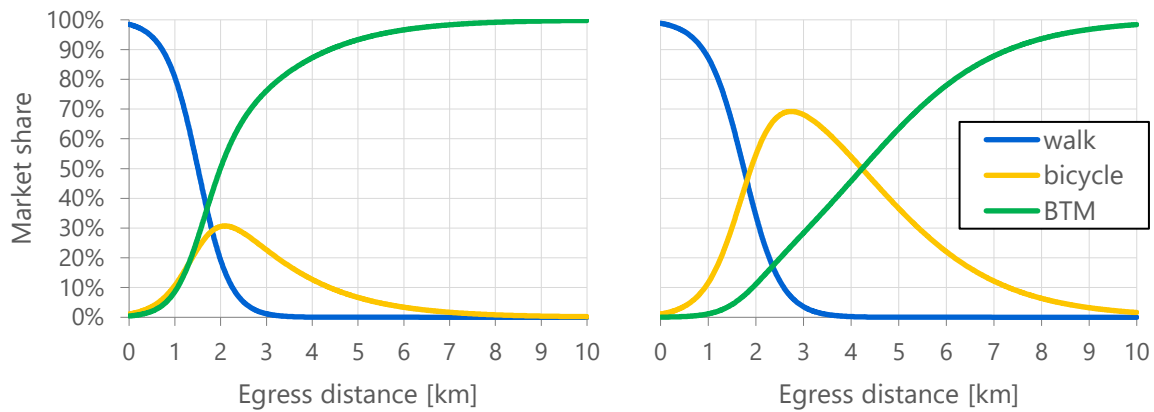

*Figure 10. Mode preferences on the egress leg, if the BTM option has 0 transfers (left) or 1 transfer (right) for reaching the destination from the train station*

## 6.4 Policy implications and recommendations

Based on the findings presented above, we outline four key recommendations for policy and practice. Firstly, improve egress options of existing stations and add new stations in urban areas, especially in areas with high job density. We see from our results that work-related trips and trips destined for urban areas show the greatest mode-shift potential and should therefore be the primary target for a shift to trains. Travellers already find trains to be more attractive in those contexts so increasing the modal split for those trips should be easier.

Secondly, increase the number of stations along corridors where most travellers would make use of a direct (non-transfer) service. Given that travel time in PT is less negative than access/egress time, adding new stations would reduce the average access distance to the nearest train station. Adding stations however needs to be balanced with ensuring direct train services for as many travellers as possible, as the benefits of shorter access distances can quickly be outweighed by imposing an additional transfer, diminishing all the benefits of the new station.

Thirdly, Intercity trains should stop more frequently. In the Netherlands, Intercity trains are the limited stopping service that tend to traverse the country and only stop in larger settlements and in the highest density areas. The new stops should be more than 5km (distance up to which people are willing to cycle to avoid a transfer) from the nearest Intercity stop and should provide direct connections to new destinations. Given the relatively low impact of public transport travel time on the overall disutility, an additional stop or two would make little impact in terms of travel time perception, while providing a substantial number of travellers with new direct travel alternatives.

Fourthly, focus on cycling infrastructure at origin stations (home-end) through improved cycle lanes and parking facilities. For distances of 1-5km, cycling is by far the most preferred mode and given its limited environmental impact and spatial footprint (Mbugua et al., 2025), it should be encouraged as much as possible. If the population density is also high beyond 5km with no or limited train stations, BTM services should also complement the cycling infrastructure for those living further away. In lower density areas, where high quality BTM cannot be justified, park-and-ride facilities can help in attracting those living further from the station.

Fifthly, focus on walkability and BTM services at destination stations (activity-end), with shared bicycles coming into play in lower density areas where high-quality BTM cannot be justified. For all destinations up to 1.5-2km, walking tends to be the most preferred, so walkable areas in the vicinity of stations should be prioritised. If there is substantial density also beyond this distance, high quality BTM (direct, dense, frequent) should be encouraged. Shared bicycles can provide a good alternative to BTM and have a substantially larger added value in lower density areas, where high-quality BTM services cannot be justified.

## 7 Conclusions

This paper analyses the preferences and choice behaviour of travellers in door-to-door trips, comparing private unimodal alternatives with a train-based multimodal option. We include a series of choices made by travellers in multimodal trips – the access mode, boarding station, alighting station and egress mode – and compare it to a unimodal trip by car, bicycle or walking. Using revealed preference data from the Dutch Mobility Panel, we estimate a nested logit model to disentangle the impact of individual trip characteristics on the journey as a whole.

Our findings show that, in as the main mode, private modes like walking, cycling and the car, are preferred to both train and bus/tram/metro (BTM). Owning a car further reduced the attractiveness of train and BTM, while walking and cycling are unaffected. Work-related trips make all other modes more attractive compared to walking. While private modes may be preferred on the whole, they do seem to be associated with a higher travel-time penalty, with public transport (PT) (train and BTM) travel time being almost 60% lower, more than the 30-50% decrease found by most studies in the field (Bovy & Hoogendoorn-Lanser, 2005; Geržinič et al., 2023; van Mil et al., 2021). The high dispreference for PT and low PT IVT perception suggests that there is a strong initial barrier to using PT, but once chosen, the time spent travelling is not perceived so negatively. The high initial penalty could be due to the multi-leg nature of train (PT) trips or a general aversion towards it. Low IVT perception on the other hand may be linked to the opportunity of using travel time productively.

On access and egress trip legs, walking tends to be the most preferred. Cycling is primarily attractive on the access leg while BTM is the mode of choice for longer egress trips, aligning with most access/egress research (Bovy & Hoogendoorn-Lanser, 2005; Geržinič et al., 2023; Halldórsdóttir et al., 2017; Keijer & Rietveld, 2000; La Paix Puello & Geurs, 2014). These preferences do not seem to be affected by either car ownership or trip purpose. The valuation of travel time is also the same as for the main mode, with BTM being equal to train IVT, whereas walking, cycling and driving are perceived equally and more negative than PT. While the ratios do not align that clearly with the results of Bovy & Hoogendoorn-Lanser (2005), the grouping of modes and how they relate to one another is identical: car, bicycle and walk are equal and more negative, while train and BTM are lower and also (almost) identical.

Lastly, transfers and urbanisation level play a big role in mode choice. Transfers are perceived highly negatively, with respondents indicating a willingness to walk, cycle or drive to a station 10min farther to avoid a transfer, or travel by BTM or train a full 25min longer. Since transfers were not modelled in more detail, this transfer penalty is the average of all transfers and includes any prospective walking, climbing up/down a set of stairs, waiting and also the potential of missing the connection and being late/having to wait for the next departure.

Urbanisation level has been used as a proxy for driving and parking difficulty, indicating that drivers especially want to avoid driving into dense urban areas, accepting an additional 28min by train to go there, or alternatively driving 11min farther by car to reach a different destination in a less urbanised area.

While great care was taken in our study, concessions had to be made in certain cases based on the available data and resources. Although multiple model specifications were tested, none of them justified the inclusion of a cost parameter. Due to the nature of ex-post price determination in RP models, the travel costs are often correlated with travel time. While this is logical and common, it causes difficulties for discrete choice models trying to disentangle the impact of individual effects. While our model does not include a direct cost parameter, it is effectively included in the time parameter.

In the modelling approach, the panel structure of the data could not be considered due to the already complex model and computational limitations. A panel mixed logit or latent class choice model could better account for such data and provide additional insights into the heterogeneity of travellers and potential traveller segments. Due to the limited occurrence of certain modes (especially non-driver car modes, like car passenger or taxi), we needed to aggregate them into five categories to enable a meaningful model estimation. Additional data on trips with these more niche modes or a more qualitative approach would provide additional insights into what makes these modes appealing, what is preventing their wider adoption and what could further encourage their use. Due to already complex structure and limited data available, we also forwent a more detailed modelling of transfers, i.e. walking time/distance, type of transfer (cross-platform, cross-station), waiting time, likelihood of making the connection etc. To better understand the penalty associated with transfers, more detailed data is needed to tackle this problem.

Additionally, we outline a few more potential research directions: firstly, as our research could not account for the impact of cost on the trip due to the high collinearity, we are unable to directly extract the willingness to pay from our study. Although our results of IVT and modal preference are solid and align with past literature, future research should work towards including travel cost in such studies as well, making the conclusions even more robust. Secondly, by including all possible modes-station combinations, we likely overestimated the scope of the consideration set of respondents. Although Arriagada et al. (2025) show that using a choice set of past observations is superior, that approach also makes strong assumptions, that anything not chosen is not even considered. Future research could adopt the approach of Hoogendoorn-Lanser & Van Nes (2004), directly asking respondents for their consideration set. Thirdly, RP datasets should be enriched with observations of shared micromobility as an egress mode. This shows to be a promising way to close the last mile (Geržinič et al., 2025; Oeschger et al., 2023), but studies have so far relied almost exclusively on stated preference data, which the hypothetical bias as its biggest downside.

# Appendix

## A. Mode aggregation

In the MPN, a total of 29 different modes can be recorded. In our dataset, 23 appear at least once as either an access, main or egress mode. To simplify the choice model estimation, we aggregate them into five mode groups. Table 4 presents the frequency with which each mode appears within the group and trip leg. For clarity's sake, if the mode does not appear at all in a specific trip leg (0%), it is not shown at all.

*Table 4. Mode aggregation and share of trips made by each mode within each group per trip leg*

| Aggregate modes | Originally coded modes | access | main | egress |
|---|---|---|---|---|
| Walking | Walking | 100.0% | 98.1% | 99.7% |
| | Other * | | 1.9% | 0.3% |
| Cycling | Bicycle, racing bicycle | 83.6% | 58.3% | 72.2% |
| | Electric bicycle, e-bike | 11.9% | 36.7% | 5.6% |
| | Public Transport rental bicycle | | 0.1% | 11.1% |
| | Folding bicycle | 3.0% | 0.4% | 11.1% |
| | Speed pedelec (max. 45 km/h, helmet required) | | 0.5% | |
| | Scooter | | 2.3% | |
| | Moped | 1.5% | 1.9% | |
| Bus/Tram/Metro | Express tram/metro | 16.7% | 38.1% | 24.5% |
| | Tram | 11.9% | 14.4% | 24.5% |
| | Bus | 71.4% | 47.5% | 51.0% |
| Train | Train | | 100.0% | |
| Car | Car as driver (gasoline, diesel or LPG) | 61.8% | 65.8% | |
| | Car as driver (electric, hydrogen or hybrid) | 8.8% | 9.8% | |
| | Car as passenger (gasoline, diesel or LPG) | 20.6% | 17.9% | 66.7% |
| | Car as passenger (electric, hydrogen or hybrid) | 8.8% | 2.6% | 33.3% |
| | Van | | 2.0% | |
| | Motorcycle | | 0.7% | |
| | Camper | | 0.1% | |
| | Handicap-accessible vehicle | | 0.7% | |
| | Taxi | | 0.4% | |
| | Touring car/bus (not Public Transport) | | 0.1% | |

\* *Trips recorded as "Other" are individually inspected (3 cases) and based on the travel distance and time, we conclude they can all be classified as walking*

## B. Nesting structures

We test a series of nesting structures to see which best captures the observed choice behaviour. We consider all train-related alternatives for potential nesting structures. Table 5 shows that the simplest structure, with a single nest with all train alternatives performs best both in terms of model fit and parsimony.

Other nesting structures are based on a shared access/egress mode/station, i.e. all alternatives that include cycling as an access mode or all alternatives egressing at the closest station etc.

We also test combinations of two of these combined, for example all alternatives that include using BTM to the third closest station or all alternatives that use walking on both the access and egress side.

*Table 5. Model outcomes for different nesting structures*

| Nesting structure | Number of nests | Final LL | BIC |
|---|---|---|---|
| Train | 1 | -3,690 | 7,534 |
| Access modes | 4 | -3,718 | 7,607 |
| Access stations | 3 | -3,716 | 7,603 |
| Egress modes | 3 | -3,715 | 7,592 |
| Egress stations | 3 | -3,718 | 7,608 |
| Access modes + stations | 12 | -3,717 | 7,674 |
| Egress modes + stations | 9 | -3,716 | 7,646 |
| Access + egress modes | 12 | -3,704 | 7,647 |
| Access + egress stations | 9 | -3,726 | 7,666 |

## C. Market share calculations

To calculate the market shares across different distance, we use the average values from our dataset. When varying the market share over egress distances, we use the average access leg and main leg distance to calculate the contribution of utility from those legs, with the utility contribution of the egress leg varying based on the distance. Equally, we use average main leg and egress leg distances when varying the access leg distance.

Next, as the used distances are Euclidian (great circle distance), we use the per-mode distance data obtained through the APIs to determine the detour factor of each mode. This is the ratio between the actual distance and the Euclidian distance between two points. To obtain travel time of each mode, we use the average speed per mode and leg, which we determine based on travel distance and travel time which in turn also come from the API data. Given that the average trip is quite long (>35km), we only consider train and car as viable main modes. The average detour factors and speeds per mode and leg are listed in Table 6.

For the context parameters, we take the average effect of car ownership interaction and trip purpose interaction. For the urbanisation level, we assume the origin is non-urban and the destination is highly urban. For the rest, we use the parameters as reported in Table 2 in Section 5.

*Table 6. Detour factors and average speeds per mode and leg*

|  | Detour factor | | | Average speed [km/h] | | |
|---|---|---|---|---|---|---|
|  | Access leg | Main leg | Egress leg | Access leg | Main leg | Egress leg |
| Average Euclidian distance [km] | 1.799 | 35.718 | 1.121 |  |  |  |
| Walking | 1.286 |  | 1.326 | 4.346 |  | 4.357 |
| Cycling | 1.424 |  | 1.520 | 16.660 |  | 15.741 |
| Car | 1.860 | 1.323 |  | 27.278 | 63.059 |  |
| BTM | 1.860 [1] |  | 1.860 [2] | 19.255 |  | 19.255 [2] |
| Train |  | 0.967 |  |  | 92.173 |  |

[1] As the API does not provide information on BTM distance, we assume the same distance as for the car
[2] Since there is no car detour factor or speed on the egress leg, we assume the same value for BTM on the egress leg as on the access leg